%% file: highpt.tex
\documentclass[twocolumn,preprintnumbers,amsmath,amssymb]{revtex4}

\usepackage{graphicx}
\usepackage{dcolumn}
\usepackage{bm}

\begin{document}

\title{Charged hadron transverse momentum distributions in Au+Au collisions \\
at $\sqrt{s_{_{NN}}} =$ 200 GeV }
\input{Phobos_authors_current}
\date{\today}

\begin{abstract}\noindent

We present transverse momentum distributions of charged hadrons 
produced in Au+Au collisions at $\sqrt{s_{_{NN}}} =$ 200 GeV. The spectra
were measured for transverse momenta $p_T$ from 0.25 to 4.5~GeV/c in a 
rapidity range of $0.2 < y_{\pi} < 1.4$. The evolution of the spectra
is studied as a function of collision centrality, from 65 to 344 
participating nucleons.  
The results  are compared to data from proton-antiproton collisions and 
Au+Au collisions at lower RHIC energies. 
We find a significant change of the spectral shape between proton-antiproton 
and peripheral Au+Au collisions. Comparing peripheral to central Au+Au 
collisions, we find that the yields at high $p_T$ exhibit 
approximate scaling with the number of participating nucleons, rather than
scaling with the number of binary collisions.
\end{abstract}

\maketitle

The yield of charged hadrons produced in collisions of gold nuclei at 
an energy of $\sqrt{s_{_{NN}}} = 200$~GeV has been measured 
with the PHOBOS detector at the Relativistic Heavy-Ion Collider (RHIC)
at Brookhaven National Laboratory.
The data are  presented as a function of collision centrality and 
transverse momentum $p_T$.
The goal of our measurements is to test our understanding of 
Quantum Chromodynamics (QCD), the fundamental theory of the strong interaction.
Using nuclear collisions, QCD can be studied in a regime of high temperature 
and energy density.  Calculations suggest \cite{qgp} that under these  conditions 
a new state of matter, the quark-gluon plasma, will be formed.

In the theoretical analysis of particle production in hadronic and nuclear collisions,
a distinction is often made between the 
relative contributions from ``hard'' parton-parton
scattering processes and ``soft'' processes. The former can be calculated 
using perturbative QCD, whereas the latter are typically treated by phenomenological models
that describe the non-perturbative sector of QCD \cite{pythia}.  
The contribution 
from hard processes is expected to grow with increasing collision energy and 
to dominate particle production at high transverse momenta.
Collisions of heavy nuclei offer ideal conditions to test our understanding 
of this picture, as ``hard'' processes are expected to scale with 
the number of binary nucleon-nucleon collisions $N_{coll}$, whereas ``soft'' 
particle production is expected to exhibit scaling with the number of participating 
nucleons $N_{part}$. 
In Glauber-model calculations, $N_{coll}$ scales approximately as 
$(N_{part})^{4/3}$. For central collisions of  Au nuclei, this leads to 
an increase of a factor of six in the ratio of $N_{coll}/N_{part}$, relative 
to proton-proton collisions. Collision centrality is therefore a 
parameter by which the relative contributions of hard and soft processes to
particle production can be varied. It was found that a description based on a combination
of hard and soft particle production is compatible with
the observed centrality dependence of the charged particle multiplicity 
at mid-rapidity in Au+Au collisions \cite{phenix_cent,phobos_cent_200}, 
although
the result can also be described by models which incorporate initial 
state parton saturation \cite{kharzeev}. A 
simultaneous study of the $p_T$ and centrality dependence of particle yields
may allow us to better discriminate between these different physical pictures.

For Au+Au collisions at RHIC energies, it has been predicted that the yield and 
momentum distribution of particles produced by hard scattering processes 
may be modified by ``jet quenching'',
{\it i.e.}\ the energy loss of high momentum partons in the dense medium
\cite{jet_quench_theory}.
This phenomenon has been proposed as a diagnostic tool for characterizing the
parton density in the initial stage of high-energy nuclear collisions. First results 
for central Au+Au collisions at $\sqrt{s_{_{NN}}} = $ 130~GeV have shown 
that particle spectra at large transverse momenta indeed change in comparison 
to p+p collisions \cite{jet_quench_phenix}. This result has been used to 
estimate the energy loss in the dense medium \cite{gyulassy_01}. However,
calculations have not yet shown a consistent interpretation of the data on spectral shapes 
and particle multiplicities. The latter seem to be overpredicted \cite{phobos200}
when the effects of jet quenching are included. 
Here we probe the scaling of particle production
over a wide range of $N_{coll}$ and  $N_{part}$ in different regimes of $p_T$ that are 
traditionally believed to be dominated by soft or hard processes.

The data were collected using the PHOBOS 
two-arm magnetic spectrometer \cite{phobos1,phobos2,pak}. 
The spectrometer arms 
are each equipped with  16 layers of Silicon sensors, providing 
charged particle tracking both outside and 
inside the 2~T field of the PHOBOS magnet. Particles within the geometrical 
acceptance region used in this analysis traverse at least 14 of the layers.
A two layer silicon vertex detector covering $|\eta| < 0.92$ and 25\% of 
the azimuthal angle provided additional information on the position
of the primary collision vertex.
In total, 135168 detector elements were read out, 
of which less than 2\% were non-functional. 

The primary event trigger was provided by two sets of 16 scintillator 
paddle counters covering pseudorapidities 
$3 < |\eta |< 4.5$.
Additional information for event selection was obtained from two 
zero-degree calorimeters which detect spectator neutrons. 
Details of the event selection and centrality determination 
can be found in \cite{phobosprl,phobos_cent_200}. 
Monte Carlo (MC) simulations of the apparatus
were based on the HIJING event generator \cite{hijing} 
and the GEANT~3.21 simulation package, folding in
the signal response for scintillator counters and silicon sensors.

For this analysis, the events were divided into six centrality classes,
based on the observed signal in the paddle counters. Given the monotonic 
relationship between the multiplicity of produced particles in the paddle
acceptance and the number of participating nucleons 
$\langle N_{part} \rangle$, the results of a  Glauber-model calculation 
were used
\cite{phobosprl,phobos_cent_200} to estimate the average number of 
participating nucleons and the number of binary collisions for each 
centrality class. The resulting values are shown in Table~I.

As the geometrical layout of the PHOBOS detector leads to an asymmetry 
in the acceptance and detection efficiency for positively and 
negatively charged particles for a given magnet polarity, 
data were taken using both polarities. 
The reproducibility of the absolute field strength was found to be 
better than 1\%, based on Hall probe measurements for each polarity
and the comparison of mass distributions for identified particles
at the two polarities.

To optimize the precision of the vertex  and track
finding, only events with a reconstructed primary vertex position 
between -10~cm $< z_{vtx} < $ 10~cm along the beam axis
were selected.
By requiring a consistent vertex position from the spectrometer and 
vertex subdetectors, in combination with the known position of the 
beam orbit, a vertex resolution better than 
0.3~mm (RMS) in the longitudinal and vertical directions, $z$ and $y$, and better 
than 0.5~mm in the horizontal direction, $x$, was achieved. 

Details of the reconstruction algorithm for particle tracks can be 
found in \cite{pbarp_200}. Track candidates were found combining a road-following algorithm
in the region outside the magnetic field and a transformation algorithm for pairs of
hits inside the field.  The distance of closest approach of each reconstructed track 
with respect to the primary vertex ($d_{vtx}$) was used to reject background 
particles from decays and secondary interactions.
The final track selection was based on the $\chi^2$ fit probability
of a full track fit, taking into account multiple scattering and energy loss.
For further analysis, particles with a rapidity of $0.2 < y_\pi < 1.4$ were used, 
assuming the pion mass for calculating $y_\pi$.

To obtain the invariant yield for charged hadrons, the observed transverse momentum 
distributions were corrected for the geometrical acceptance of the detector, the 
efficiency of the tracking algorithm and the distortion due to binning and momentum
resolution. For the acceptance and efficiency corrections, the correction factors 
in $p_T$ and $y$ vary as a function of vertex position and are different for the two 
charge signs in a given field polarity. 
The correction factors were obtained in vertex bins of $\pm 2.5$~cm size by embedding 
the same MC particle tracks into ``empty'' events and data events.
The events were reconstructed with the standard reconstruction chain 
with and without the additional embedded tracks. From this comparison, a correction
factor as a function of $p_T$ and multiplicity was extracted. Over the centrality
range covered in this analysis, the reconstruction efficiency decreased by approximately 
10\%.

MC-simulations based on HIJING and GEANT were used
to obtain correction factors for the contribution of secondary particles to the charged particle spectra. 
The acceptance for secondary and feeddown particles is limited
to those produced within 10~cm radial distance from the primary collision vertex,
as accepted tracks were required to have at least one hit in 
the first two layers of the spectrometer.
The background contamination is further reduced by requiring the particle 
tracks to have $d_{vtx} < 3.5$~mm. 
We found that the contamination varies from less than 4\% at $p_T \approx 0.3$~GeV/c to
less than 2\% at $p_T > 2.5$~GeV/c. The corresponding correction was applied independent of collision
centrality.

The contamination by wrongly reconstructed (``ghost'') tracks was studied using two methods. The reconstruction
of HIJING events showed a rate of ghost tracks after all quality cuts of less than 4\% at $p_T \approx 0.5$~GeV/c
and less than 2\% at  $p_T \approx 2.5$~GeV/c. At higher $p_T$, the available MC statistics did not allow a 
precise determination of the ratio of ghost tracks to found tracks. However, MC simulations
 showed that most
ghost tracks arise from a wrong match of track segments inside and outside the magnetic field. The rate of 
accidental matches was tested by swapping straight and curved track segments between the two spectrometer
arms. We found a contribution of accidental matches of less than 1\% in the $p_T > 3$~GeV/c region.

Finally, the spectra were corrected for the effects of the finite bin-width and the momentum resolution
using an iterative procedure. The resolution was determined using track embedding and the 
full reconstruction and fitting procedure.  For embedded MC tracks we found a momentum resolution of 
$\Delta p/p = 1 - 9\%$ for the momentum range of 0.3 to 10 GeV/c. 
The corresponding correction was performed for each centrality bin 
separately, taking into account the non-Gaussian tails in the $p_T$-resolution function. 
At $p_T \approx 4 $~GeV/c, the correction varies from a factor of 0.92 for the most peripheral events 
to 0.89 for the most central events. 

\begin{figure}[t]
\includegraphics[width=8cm]{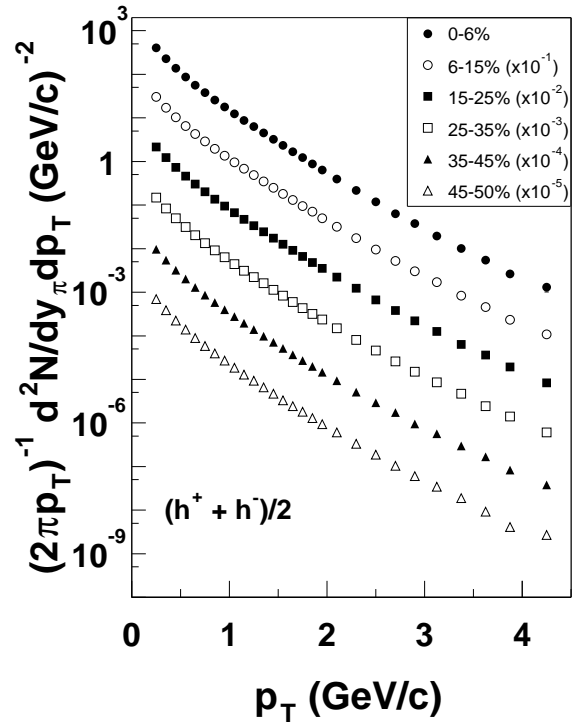}
\caption{ \label{ref_SpectraAllCent} 
Invariant yields for charged hadrons as a function of $p_T$ for 6 
centrality bins. For clarity, consecutive bins are scaled by factors of 10.
Statistical and systematic uncertainties are smaller than the symbol size.}
\end{figure}

\begin{table}[htbp]
\begin{center}
\begin{tabular}{|c|c|c|c|}
\hline Centrality & N$_{events}$ &  
$ \left<N_{part}\right>$ &  $ \left< N_{coll}\right>$ \\
\hline
\hline
45-50\% & 593942  &  65   $\pm 4$   & 107  \\
35-45\% & 1196402  &  93   $\pm 5$  & 175  \\
25-35\% & 1192357  &  138  $\pm 6$ & 300  \\
15-25\% & 1193373  &  200  $\pm 8$ & 500  \\
6-15\%  & 1057007  &  276  $\pm 9$ & 780  \\
0-6\%   &  604555  &  344  $\pm 11$ & 1050 \\
\hline
\end{tabular}
\caption{
\label{table1}
Details of the centrality bins used. The estimated uncertainty in 
$\langle N_{coll} \rangle$ ranges from
15\% to 10\% going from the most peripheral to the most central bin.  }

\end{center}  
\end{table} 

\begin{figure}[t]
\includegraphics[width=7.5cm]{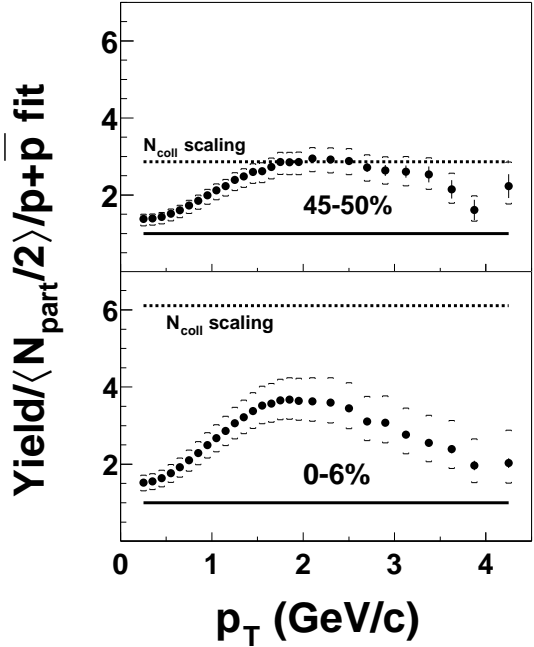}
\caption{ \label{ref_SpectraRatioPbarP}
Ratio of the yield of charged hadrons as a function of $p_T$ for
the most peripheral bin ($\langle N_{part} \rangle = 65 \pm 4$, upper plot)
and the most central bin ($\langle N_{part} \rangle = 344 \pm 11$, lower plot) to a fit 
of proton-antiproton data (see text) scaled by $\langle N_{part}/2\rangle$.
The dashed (solid) line shows the expection of $N_{coll}$ ($N_{part}$) scaling 
relative to $p+\bar{p}$ collisions. 
The brackets show the systematic uncertainty of the Au+Au data.}
\end{figure}

\begin{figure}[t]
\includegraphics[width=7.5cm]{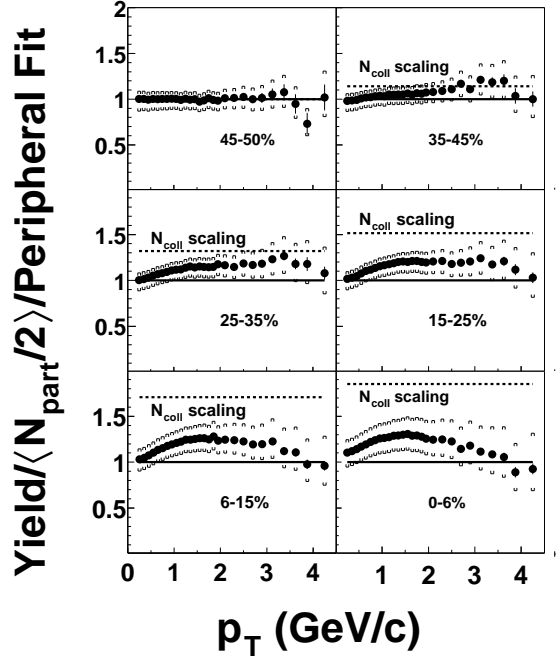}
\caption{ \label{ref_RatioPeripheral}
Charged hadron yield in Au+Au in six centrality bins, divided by
a fit to the most peripheral bin and by 
$\langle N_{part} /2 \rangle$.
The dashed (solid) line shows the expectation for $N_{coll}$ ($N_{part}$) scaling relative
to peripheral collisions. The brackets show the systematic 
uncertainty in the Au+Au data.}
\end{figure}

In Fig.~\ref{ref_SpectraAllCent} we show the invariant yield
of charged hadrons as a function of transverse momentum, obtained by 
averaging the yields of positive and negative hadrons. 
Data are shown for 6 centrality bins, 
ranging from $\langle N_{part}\rangle = 65$ to $\langle N_{part}\rangle = 344$. 
The integrated yields, when scaled by $\langle N_{part}\rangle$, increase by 15\% over the centrality range, 
consistent with the centrality evolution 
of the mid-rapidity particle density presented in \cite{phobos_cent_200}.

In Fig.~\ref{ref_SpectraRatioPbarP} we show
the ratios of data for the most peripheral bin ($\langle N_{part} \rangle = 65 \pm 4)$ and the most central bin
($\langle N_{part} \rangle = 344 \pm 11)$ to a fit of 
the measured $p_T$-distribution in proton-antiproton collisions at the same energy \cite{ua1_pbarp}. In
both cases, the Au+Au data were divided by the respective value of $\langle N_{part}/2 \rangle$. 
The brackets indicate the systematic uncertainty in the Au+Au data (90\% C.L.). The largest contribution
to the systematic uncertainty are the overall tracking efficiency, independent of $p_T$, and the 
$p_T$ dependent momentum resolution and binning correction. The small contamination by 
secondary particles and feeddown particles, due to the proximity of the tracking detectors to the 
collision vertex, also ensures that the uncertainty in the corresponding correction factors is small.
Similarly, the high granularity and resolution of the tracking planes leads to small uncertainty
in the rate of `ghost' tracks.

As has been shown previously \cite{phobos200,phobosprl}, the yield per participant pair 
in Au+Au collisions 
at these centralities is significantly larger than in proton-antiproton collisions at the same 
energy.  We also observe that in peripheral Au+Au collisions with $\langle N_{part} \rangle = 65$, corresponding to an impact parameter $b \approx 10$~fm, 
the spectral shape is already strongly modified from that in $p+\bar{p}$ collisions. 
It is worth noting that the ratio $\frac{\langle N_{coll}\rangle}{\langle N_{part} / 2\rangle}$ 
increases by a factor of almost three from $p+\bar{p}$ to 
the most peripheral Au+Au collisions studied here. For 
the highest $p_T$, the yield for central events is significantly smaller than expectations based 
on $N_{coll}$-scaling.

The detailed evolution of the spectra from peripheral to central events is shown in 
Fig.~\ref{ref_RatioPeripheral},
for the six centrality bins. Each spectrum has been been divided by ${\langle N_{part} / 2\rangle}$
and a fit to the data for the most peripheral bin. 
$N_{part}$-scaling would correspond to the ratio being constant at one (solid line), whereas
$N_{coll}$-scaling, shown by the dashed line, would lead to an increase by a factor of two from 
peripheral to central events. Again, the brackets show the systematic uncertainty (90\% C.L.).
It is remarkable that the change in spectral shape over this range of centralities is 
small compared to that between $p+\bar{p}$ collisions and peripheral events
shown in the top panel of Fig.~\ref{ref_SpectraRatioPbarP}. In particular at $p_T > 3$ GeV/c, 
the data scale with the number of participants to very good approximation.

\begin{figure}[t]
\includegraphics[width=7.5cm]{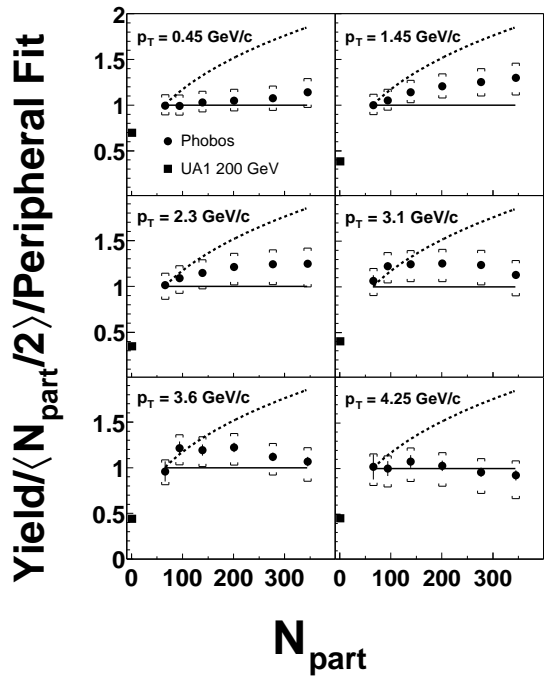}
\caption{ \label{ref_YieldCent}
Charged hadron yields per participant pair in 6 different 
transverse momentum bins, plotted as a function of $N_{part}$.
The data are normalized to the yield in the most peripheral centrality bin. 
The dashed (solid) line shows the expectation for $N_{coll}$ ($N_{part}$) scaling from
peripheral to central collisions. The brackets indicate the systematic 
uncertainty for the centrality evolution of this ratio (90\% C.L.).}
\end{figure}

The scaling behaviour in various regions of $p_T$ is further illustrated in Fig.~\ref{ref_YieldCent}. 
Here we show the yield per participant pair in six bins of $p_T$ between 0.45 and 4.25 GeV/c,
normalized to the yield per participant pair in the most peripheral bin. The expectation for $N_{coll}$-scaling 
relative to the most peripheral bin is again shown as a dashed line.
The observed evolution can be contrasted with the expectation that particle production should be characterized
by a change from  $\langle N_{part}\rangle$ scaling at low $p_T$ to $\langle N_{coll}\rangle$ scaling
at high $p_T$, with $\langle N_{part}\rangle$ increasing from  65 to 344 and $\langle N_{coll} 
\rangle$ from 107 to 1050 
between the most peripheral collisions and the most central collisions studied here. No corresponding
increase in particle production per participant at $p_T = 3$~GeV/c and above is observed. Rather,
the yields in this region scale approximately with the number of participating nucleons.

The failure of binary collision scaling at intermediate and large $p_T$ was previously 
observed in Au+Au collisions at $\sqrt{s_{_{NN}}} = $ 130~GeV \cite{jet_quench_phenix}. 
This effect has been interpreted as the result of the energy loss of jets in the dense 
medium, leading to a suppression of leading hadrons relative to the expected scaling.
The observed particle spectra are believed to depend on a complex interplay of initial 
and final state effects in the production and fragmentation of high $p_T$ partons,
including initial state multiple scattering, nuclear shadowing and parton energy loss.
Our observations show that, surprisingly, the combination of all effects leads to an apparently very 
simple centrality scaling at large $p_T$, providing a challenge to theoretical descriptions. 
That this is not accidental is suggested by the agreement with earlier measurements at 
$\sqrt{s_{_{NN}}} = $ 130~GeV. The lower energy data \cite{star_highpt_npart,phenix_highpt_npart}, 
when compared over the same 
centrality range, show very similar centrality 
scaling, even though the invariant yield at high $p_T$ increases much more rapidly with increasing 
beam energy than the overall particle yield. 
It has recently been argued that the observed scaling could be naturally explained in a 
model assuming the dominance of surface emission of high-$p_T$ hadrons \cite{mueller_8_02}.
However, the approximate participant scaling has also been explained in the context of 
initial state saturation models \cite{kharzeev_10_02}. 
Experimentally, upcoming studies of d+Au collisions at RHIC will provide further insight 
into the modification of particle spectra in a nuclear environment.

This work was partially supported by U.S. DOE grants DE-AC02-98CH10886,
DE-FG02-93ER40802, DE-FC02-94ER40818, DE-FG02-94ER40865, DE-FG02-99ER41099, and
W-31-109-ENG-38 as well as NSF grants 9603486, 9722606 and 0072204.  The Polish
group was partially supported by KBN grant 2-P03B-10323.  The NCU group was
partially supported by NSC of Taiwan under contract NSC 89-2112-M-008-024.

\end{document}

%% file: Phobos_authors_current.tex
\author{
%
%
B.B.Back$^1$,
M.D.Baker$^2$,
D.S.Barton$^2$,
R.R.Betts$^6$,
M.Ballintijn$^4$,
A.A.Bickley$^7$,
R.Bindel$^7$,
A.Budzanowski$^3$,
W.Busza$^4$,
A.Carroll$^2$,
M.P.Decowski$^4$,
E.Garc\'{\i}a$^6$,
N.George$^{1,2}$,
K.Gulbrandsen$^4$,
S.Gushue$^2$,
C.Halliwell$^6$,
J.Hamblen$^8$,
G.A.Heintzelman$^2$,
C.Henderson$^4$,
D.J.Hofman$^6$,
R.S.Hollis$^6$,
R.Ho\l y\'{n}ski$^3$,
B.Holzman$^2$,
A.Iordanova$^6$,
E.Johnson$^8$,
J.L.Kane$^4$,
J.Katzy$^{4,6}$,
N.Khan$^8$,
W.Kucewicz$^6$,
P.Kulinich$^4$,
C.M.Kuo$^5$,
W.T.Lin$^5$,
J.W.Lee$^4$,
S.Manly$^8$,
D.McLeod$^6$,
A.C.Mignerey$^7$,
R.Nouicer$^6$,
A.Olszewski$^3$,
R.Pak$^2$,
I.C.Park$^8$,
H.Pernegger$^4$,
C.Reed$^4$,
L.P.Remsberg$^2$,
M.Reuter$^6$,
C.Roland$^4$,
G.Roland$^4$,
L.Rosenberg$^4$,
J.Sagerer$^6$,
P.Sarin$^4$,
P.Sawicki$^3$,
W.Skulski$^8$,
S.G.Steadman$^4$,
P.Steinberg$^2$,
G.S.F.Stephans$^4$,
A.Sukhanov$^2$,
J.-L.Tang$^5$,
R.Teng$^8$,
A.Trzupek$^3$,
C.Vale$^4$,
G.J.van~Nieuwenhuizen$^4$,
R.Verdier$^4$,
G.I.Veres$^4$,
B.Wadsworth$^4$,
F.L.H.Wolfs$^8$,
B.Wosiek$^3$,
K.Wo\'{z}niak$^3$,
A.H.Wuosmaa$^1$,
B.Wys\l ouch$^4$\\
\vspace{3mm}
\small
%
%
%
%
$^1$~Argonne National Laboratory, Argonne, IL 60439-4843, USA\\
$^2$~Brookhaven National Laboratory, Upton, NY 11973-5000, USA\\
$^3$~Institute of Nuclear Physics, Krak\'{o}w, Poland\\
$^4$~Massachusetts Institute of Technology, Cambridge, MA 02139-4307, USA\\
$^5$~National Central University, Chung-Li, Taiwan\\
$^6$~University of Illinois at Chicago, Chicago, IL 60607-7059, USA\\
$^7$~University of Maryland, College Park, MD 20742, USA\\
$^8$~University of Rochester, Rochester, NY 14627, USA\\
}